\documentclass[aps,prl,twocolumn,groupedaddress,floatfix,superscriptaddress]{revtex4}
\usepackage{graphicx}

\begin{document}

\newcommand{\gc}{\ensuremath{\Gamma_{\mathrm c}}}
\newcommand{\tc}{\ensuremath{T_{\mathrm c}}}
\newcommand{\ga}{\ensuremath{\Gamma_{\mathrm a}}}
\newcommand{\tint}{\ensuremath{t_{\rm int}}}
\newcommand{\us}{\ensuremath{\mu{\rm s}}}
\newcommand{\um}{\ensuremath{\mu{\rm m}}}
\newcommand{\uw}{\ensuremath{\mu{\rm W}}}
\newcommand{\ba}{\mbox{\ensuremath{^{138}{\rm Ba}}}}
\newcommand{\transitiontriplet}{\mbox{\ensuremath{^1{\rm S}_0\leftrightarrow\,\!^3{\rm P}_1}}}
\newcommand{\transitionsinglet}{\mbox{\ensuremath{^1{\rm S}_0\leftrightarrow\,\!^1{\rm P}_1}}}
\newcommand{\be}{\begin{equation}}
\newcommand{\ee}{\end{equation}}
\newcommand{\bd}{\begin{displaymath}}
\newcommand{\ed}{\end{displaymath}}
\newcommand{\bfig}{\begin{figure}}
\newcommand{\efig}{\end{figure}}
\newcommand{\f}{\ensuremath{\mathcal{F}}}
\newcommand{\fsr}{\ensuremath{\mathrm FSR}}
\newcommand{\gtwo}{\ensuremath{g^{(2)}}}
\newcommand{\sno}{791 nm}
\newcommand{\fft}{553 nm}
\newcommand{\Nth}{\ensuremath{N_{\rm th}}}
\newcommand{\nth}{\ensuremath{n_{\rm th}}}
\newcommand{\nex}{\ensuremath{n_{\rm ex}}}
\newcommand{\Nex}{\ensuremath{N_{\rm ex}}}
\newcommand{\Neff}{\ensuremath{N_{\rm eff}}}
\newcommand{\Neffnot}{\ensuremath{N_{\rm eff}^{0}}}
\newcommand{\sngt}{\ensuremath{\sqrt{n+1}\ g \tint}}
\newcommand{\mhz}{\ensuremath{{\rm MHz}}}
\newcommand{\khz}{\ensuremath{{\rm kHz}}}
\newcommand{\tem}{\ensuremath{{\rm TEM}_{00}}}
\newcommand{\ocav}{\ensuremath{\omega_{\rm cav}}}

\title{Observation of Multiple Thresholds in the Many-Atom Cavity QED
Microlaser}

\author{C.~Fang-Yen} 

\affiliation{G.~R.~Harrison Spectroscopy Laboratory, Massachusetts
Institute of Technology, Cambridge, MA 02139.}

\author{C.~C.~Yu}
\affiliation{G.~R.~Harrison Spectroscopy Laboratory, Massachusetts
Institute of Technology, Cambridge, MA 02139.}

\author{S.~Ha}
\altaffiliation{Now at Department of Physics, University of Wisconsin, Madison, Wisconsin}
\affiliation{G.~R.~Harrison Spectroscopy Laboratory, Massachusetts
Institute of Technology, Cambridge, MA 02139.}

\author{W.~Choi}
\affiliation{Department of Physics, Seoul National University, Seoul, Korea}

\author{K.~An}
\affiliation{Department of Physics, Seoul National University, Seoul, Korea}

\author{R.~R.~Dasari}
\affiliation{G.~R.~Harrison Spectroscopy Laboratory, Massachusetts
Institute of Technology, Cambridge, MA 02139.}

\author{M.~S.~Feld}
\email{msfeld@mit.edu}
\affiliation{G.~R.~Harrison Spectroscopy Laboratory, Massachusetts
Institute of Technology, Cambridge, MA 02139.}


\date{\today}

\begin{abstract}
We report the observation of multiple laser thresholds in the
many-atom cavity QED microlaser.  Traveling-wave coupling and a
supersonic atom beam are used to create a well-defined atom-cavity
interaction.  Multiple thresholds are observed as jumps in photon
number due to oscillatory gain.  
Although the number of intra-cavity atoms is large, up to $N \sim
10^3$, the dynamics of the microlaser agree with a single atom theory.
This agreement is supported by quantum trajectory simulations of a
many-atom microlaser and a semiclassical microlaser theory.
We discuss the relation of the microlaser with the micromaser and
conventional lasers.



\end{abstract}


\maketitle


The most fundamental model of light-matter interaction at the atomic
level consists of a two-level atom coupled to a single mode of the
electromagnetic field, e.\,g.\ in an optical resonator.  For an
atom-cavity coupling strength greater than the decay rates of atom and
cavity ($g \gg \gc, \ga$), an excited atom exchanges energy coherently
with the cavity (Rabi oscillation).  \cite{J-C-IEEE}

Atom-resonator interaction is also the domain of laser physics;
therefore it may be somewhat surprising that most descriptions of
laser operation use an {\it incoherent} model involving population
densities and Einstein $A$ and $B$ coefficients.  Such a model applies
due to gain medium broadening, laser field nonuniformity, and other
statistical effects \cite{scully-zubairy-quantum-optics}.

The microlaser is to our knowledge the first laser in which coherent
Rabi oscillation is explicitly reflected in the dynamics of the laser.
A controlled atom-cavity interaction and absence of strong statistical
averaging leads to behavior foreign to conventional lasers.  In this
paper we report the most dramatic of these effects, the existence of
multiple laser thresholds.

The microlaser is the optical analogue of the micromaser
\cite{Meschede-PRL85}, in which a similar bistable behavior has been
observed for a single atom in the cavity \cite{benson-prl94}.  The
most significant differences between the two experiments are: (i) a
large number of atoms is present, (ii) optical frequencies allow
direct detection of generated light, and (iii) the number of thermal
photons at optical frequencies is negligible.

An earlier version of the microlaser experiment \cite{An-PRL94} was
shown to exhibit laser action with an intracavity atom number $N$ on
the order of 1.  In the current setup, $N \gg 1$; remarkably, the
many-atom system exhibits very similar behavior to that predicted from
a single atom theory.  This is supported by quantum trajectory
simulations and may be explained by a semiclassical theory
(\cite{kwan-sc-theory}, \cite{sc-theory}).


\bfig
\centerline{\resizebox{3.0in}{!}{\includegraphics{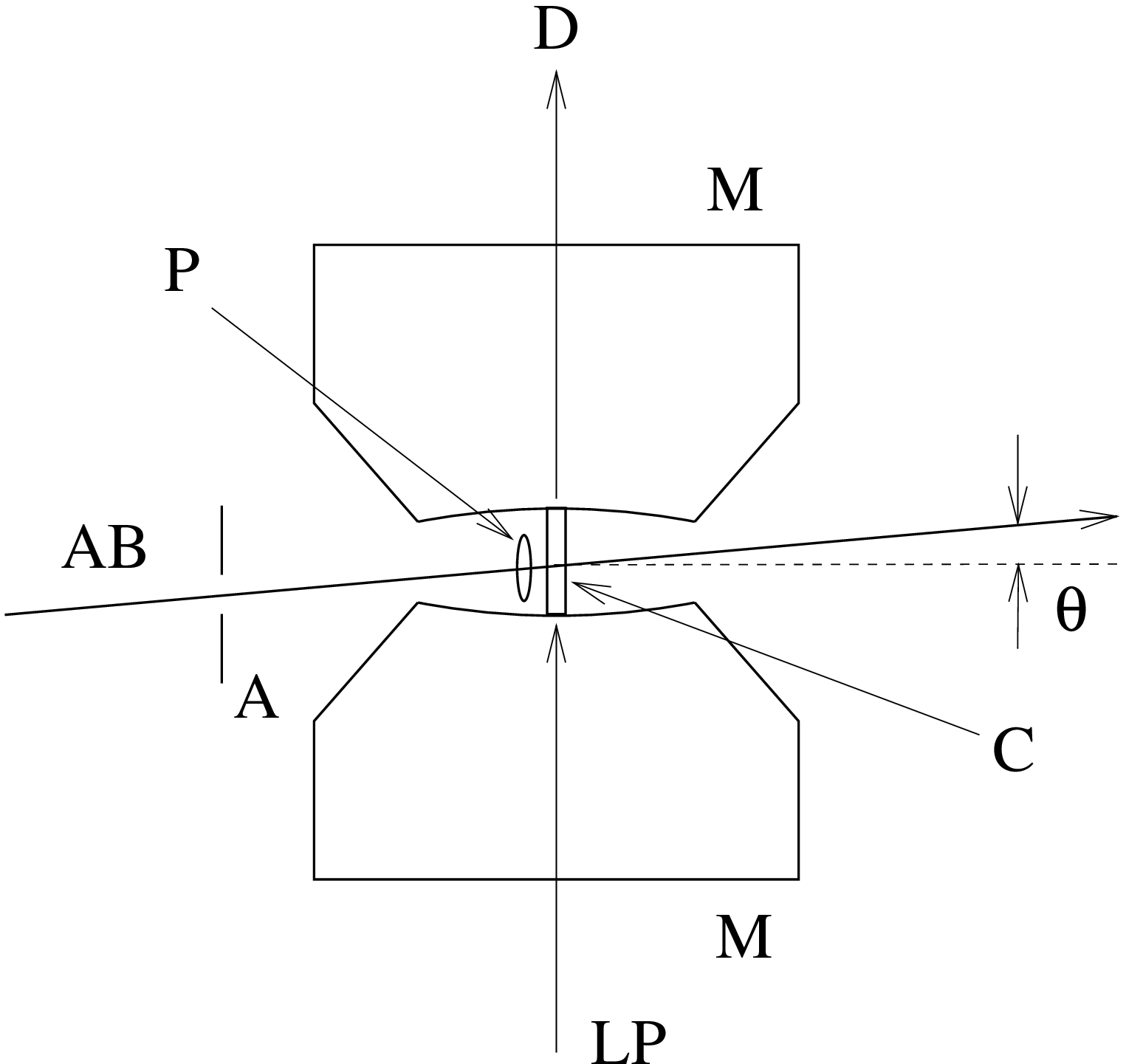}}}
\caption{Schematic of cavity QED microlaser.  M: cavity mirrors, C:
cavity mode, AB: atomic beam from oven, A: collimating aperture, P:
pump field (into page), $\theta$: cavity tilt angle, LP: locking probe
beam.  Dashed line is normal to cavity axis.}
\label{fig-microlaser}
\efig

The experiment is illustrated in Fig.~\ref{fig-microlaser}.  A
supersonic beam of \ba\ atoms passes through the \tem\ mode of a
high-finesse optical cavity (symmetric near-planar cavity, mirror
separation $L \approx 0.94$ mm, mirror radius of curvature $r_0 = 10$
cm, finesse $F \approx 9.0 \times 10^5$.)  The mirror separation and
finesse were determined by measurement of transverse mode spacing and
cavity ringdown decay time.  The background pressure in the vacuum
chamber is less than $10^{-6}$ torr.  Prior to entering the cavity
mode, each atom is excited by a cw pump laser (Coherent 899-21
Ti:Sapphire ring laser) locked to the \transitiontriplet\ transition
($\lambda = 791.1$ nm, linewidth $\ga \approx 50$ kHz) as described in
\cite{an-apl95}.  The pump beam is focused by a cylindrical lens to a
roughly $50 \um\ \times 500 \um$ elliptical Gaussian beam centered
approximately $150\ \um$ from the cavity axis.
This distance between pump and cavity mode assures that the pump beam
does not overlap the cavity, while minimizing the effect of atomic
decay between the pump and cavity.  The total decay rate from the
$^3P_1$ excited state corresponds to a decay length $\approx 1.3$ mm
for an atom velocity of 815 m/s.
The cavity spacing is adjusted by a piezoelectric modulator to be
close to resonance with the \transitiontriplet\ transition by
monitoring of a $\lambda = 791.1$ nm probe beam through the cavity.
The pump beam polarization direction is set perpendicular to the
cavity axis to ensure maximum coupling.  A magnetic field of $\approx
150$ gauss oriented parallel to the pump beam polarization ensures
that only $m=0 \leftrightarrow m=0$ transitions occur.

To measure the state of atoms which have passed through the pump beam,
we monitor fluorescence of the atoms in the focused beam of a dye
laser tuned to the \transitionsinglet, $\lambda = 553.5$ nm
transition.  The atoms were found to follow an adiabatic inversion
process due to a Doppler frequency chirp from pump beam defocus.  The
effect is similar to that described in \cite{kroon-pra85}.  Incomplete
inversion of the atoms by the pump beam is accounted for by defining
an effective atom number $N_{\rm eff} \equiv N (\rho_{ee} -
\rho_{gg})$, where $\rho_{ee}$, $\rho_{gg}$ are the excited and ground
state populations of the atoms entering the cavity.  For the data
presented here $\rho_{ee} - \rho_{gg} \approx 0.70$.

The atom-cavity interaction time $\tint = \sqrt{\pi} w_m/v_0 \approx
0.10\: \us$, with $w_m \approx 41\: \um$ the mode waist of the cavity and
central atom velocity $v_0$.

In order to observe multiple thresholds, a well-defined atom-cavity
interaction is essential, requiring a redesign of the earlier system
\cite{An-PRL94}.  

Uniform atom-cavity interaction is achieved via control of atom-cavity
coupling $g$ and interaction time \tint.  The rapid spatial variation
in coupling due to cavity standing wave is eliminated by introducing a
small tilt of the atomic beam relative to the normal to the cavity
axis (cf.\ \cite{An-OL97b}).  In this configuration the microlaser
exhibits two Doppler-shifted resonances approximately centered at
cavity frequencies $\omega_a \pm k v_0 \theta$ where $\omega_a$ is the
atom resonance frequency.

The peak traveling-wave atom-cavity coupling is given by $g_0 = (\mu
/2 \hbar) \sqrt{ 2 \pi \hbar \omega_a / V} \approx 190$ kHz where $V =
\pi L w_0^2/4$ is the cavity mode volume and $\mu$ is the dipole
matrix element.  Note that this expression for $g_0$ reflects a factor
of $2$ reduction in the peak coupling relative to the standing wave
value.

The atom beam is restricted to the center of the Gaussian mode by a
$25 \um \times 250 \um$ rectangular aperture located approximately 3
mm from the cavity axis.  The aperture is translated and rotated in
order to overlap the cavity axis as closely as possible.  The
resulting total atom-to-atom variation in peak coupling $g$ is
estimated to be $\approx 10\%$.

To ensure uniform atom-cavity interaction times, a supersonic atomic
beam with narrow velocity distribution is used.  The beam oven is
similar to that of Ref.~\cite{thomas-ol89} and consists of a
barium-filled resistively heated tantalum tube with a 250 micron
diameter nozzle.  The distance between the oven and cavity is 45 cm.
The longitudinal velocity distribution was measured via a Doppler
fluorescence measurement to be maximum at $v_0 \approx 815$ m/s and
have a width of $\Delta v_{\rm FWHM} \approx 100$ m/s.  A total FWHM
variation in $g \tint\ $ of approximately 15\% is achieved.  Atomic
density in the cavity is modulated by a translating knife edge
approximately halfway between the oven and beam aperture.  

A heuristic rate equation for the microlaser may be obtained by
equating the photon gain and loss per atom transit:
\be P_{\rm emit}(n) = \frac{\gc \tint n}{\Neff}
\label{eq-rate-equation}
\ee
where
\begin{eqnarray*}
 P_{\rm emit}(n) = \int dg f(g)\int d\tint
h(\tint) \sin^2 (\sqrt{n+1}\ g \tint) 
\end{eqnarray*}
\noindent 
is the average emission probability for an atom and $n$ is the number
of photons in the cavity.  The integrals over $g$ and $\tint$ with
respective weighting functions $f(g)$ and $h(\tint)$ reflect the
distributions in these values.  (An analogous stochastic averaging is
justified in the appendix of \cite{Filipowicz-PRA86}.)


\bfig
\centerline{\resizebox{3.5 in}{!}{\includegraphics{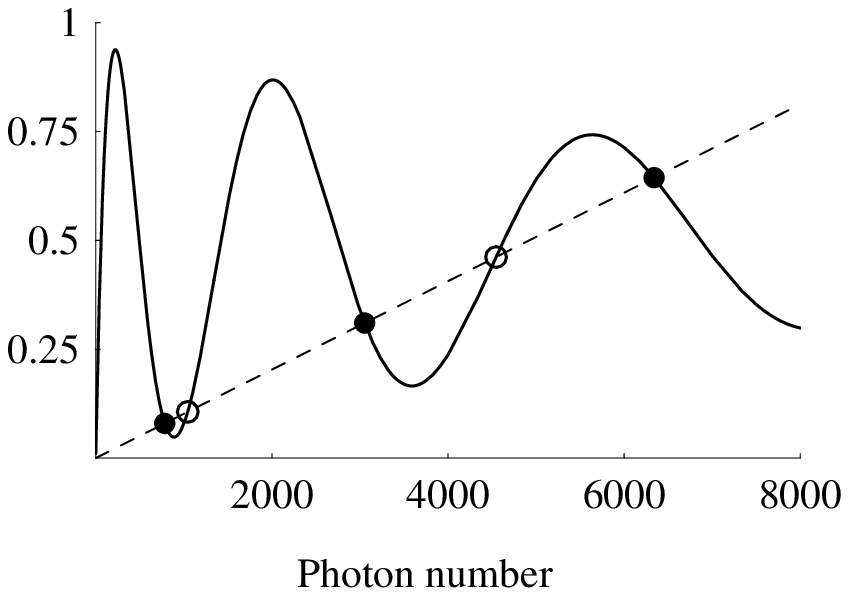}}}
\caption{Gain (solid line) and loss (dashed line) per atom transit in
the rate equation analysis (left and right sides of
Eq.~\ref{eq-rate-equation}, respectively) including effects of
velocity distribution, nonuniform coupling, and detuning, with
$\Neff=1000$.  Closed circles: stable solutions.  Open circles: unstable
solutions.}
\label{fig-gain-loss-rate-equation}
\efig

The rate equation is expressed graphically in
Fig.~\ref{fig-gain-loss-rate-equation}.  Damping in the oscillatory
gain results from averaging over $g$ and $\tint$.  A solution $n'$ of
Eqn.~\ref{eq-rate-equation} is stable if 
${\partial \over {\partial n}} (P_{\rm emit}(n)-\gc \tint n/\Neff)|_{n'} < 0$. 
For sufficiently large atom number there
exists more than one stable solution.

We now describe two experiments to observe multiple thresholds, the
transitions between solutions of Eqn.~\ref{eq-rate-equation}.

In the first experiment, the microlaser cavity was locked on resonance
($\omega_{\rm cav} = \omega_a + k v_0 \theta$) and the microlaser
emission measured as a function of atom density.  The cavity was
chopped between data collection and locking on resonance with the
atoms.  Locking was performed with a 791 nm probe beam tuned to
$\omega_a + k v_0 \theta$ via acousto-optic modulators and aligned
with the \tem\ mode of the cavity.  The cavity transmission was
monitored by an avalanche photodiode (APD) and a feedback loop
controlled the cavity PZT voltage to hold the cavity at resonance.
Photons exiting the cavity were incident on a cooled photomultiplier
tube and pulses are recorded by a photon counter.

We collected two sets of data: (i) in which the cavity photon number
was allowed to develop from an initially empty cavity ($n=0$), and
(ii) in which the cavity was ``seeded'' with a large ($n \sim
10^{25}$) number of photons by a laser pulse provided by the cavity
locking probe beam.  In both cases the microlaser emission was
observed for 100 ms.


\bfig
\centerline{\resizebox{3.2 in}{!}{\includegraphics{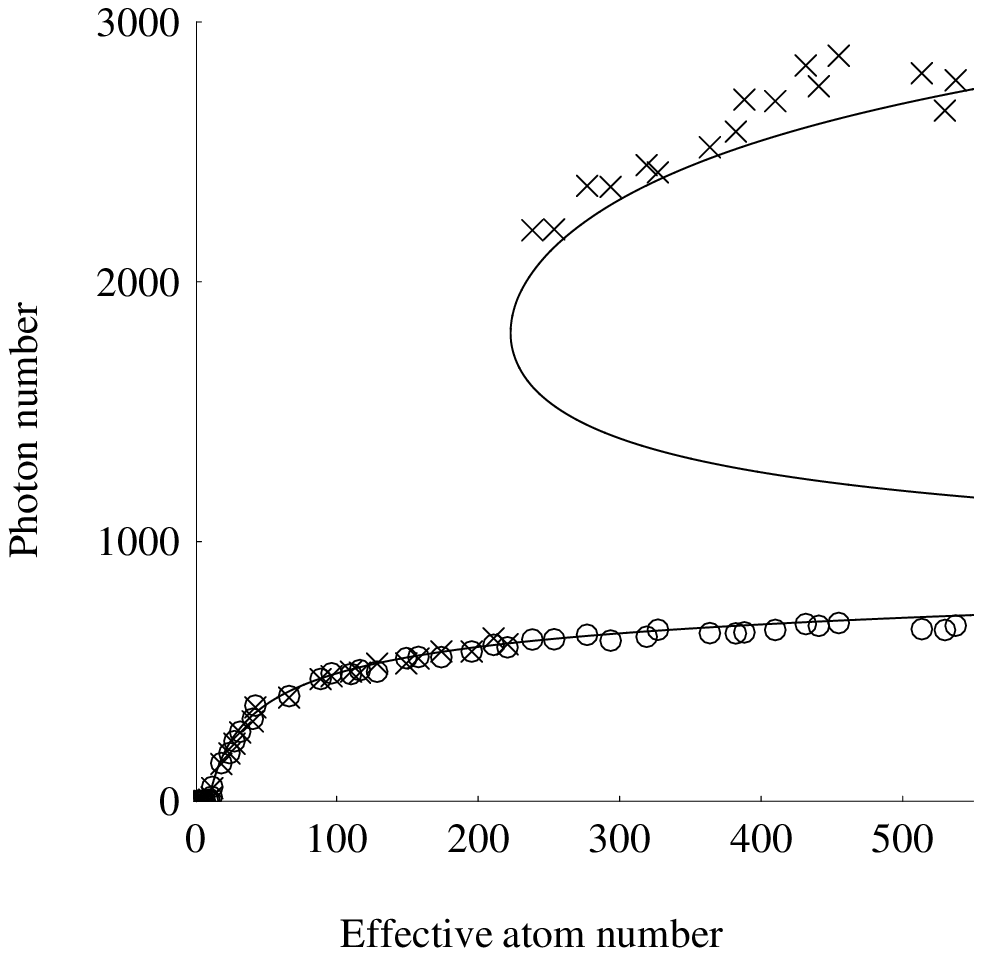}}}
\caption{Photon number $n$ vs.\ effective atom number \Neff\ for
cavity locking experiment.  Circles ($\circ$): ``unseeded'' data;
Crosses ($\times$): ``seeded'' data.  Solid line: Rate equation model,
solution of Eqn.~\ref{eq-rate-equation} for experimental parameters.}
\label{fig-cavlock}
\efig

Results from the cavity locking experiment are shown in
Fig.~\ref{fig-cavlock}.  The solid line represents the solution to
Eqn.~\ref{eq-rate-equation} and is composed of two branches for the
range of $\Neff$ shown.  The first (lower) and second (upper) branches
correspond to the first and second Rabi oscillations of
Fig.~\ref{fig-gain-loss-rate-equation}.  The lower half of the second
branch (the portion of the line with negative slope in this figure)
is unstable.

Intracavity atom number was determined by measuring fluorescence at
\transitionsinglet, $\lambda = 553.5$ nm with an imaging system and
CCD camera.
Uncertainties in cavity transmission, optical reflection coefficients,
and CCD quantum efficiency make it difficult to measure the absolute
density accurately.  A fit to the experimental data was used to obtain
a scale factor.  A fitting parameter is close to the value expected
from fluorescence measurements.  The number of photons in the cavity
is similarly estimated by considering the cavity parameters, losses in
the optics, and detector quantum efficiency.  In both cases the
fitting parameters were within $\pm 50\%$ of expected values.


The unseeded results are in good agreement with the first solution
branch.  The seeded results agree with the first branch for densities
up to the onset of bistability at $\Neff \approx 240$; above this
point the results agree with the second branch.

A calculation for steady-state average photon number using single-atom
micromaser theory of Filipowicz {\it et al} \cite{Filipowicz-PRA86}
predicts a sharp transition between the first and second branches in
at $\Neffnot \approx 415$.  No such transition was observed,
apparently due to very long metastable lifetimes in the picture of the
Fokker-Planck analysis of \cite{Filipowicz-PRA86}, due to the large
photon number.

In the second experiment, the cavity detuning $\Delta_{\rm cav} =
\omega_{\rm cav} - \omega_a$ is varied for constant atom density.
Data for the detuning curves for a range of atom densities are shown
in Fig.~\ref{fig-lineshapes}.  The solid and dashed lines represent
scans in the positive and negative detuning directions, respectively.

Two-peaked structure is due to two traveling-wave resonances of the
tilted cavity.  The Doppler splitting was measured to be about $27.1\
\mhz$, corresponding to a cavity tilt angle $\theta \approx 13$ mrad.


\begin{figure}
\centerline{\resizebox{!}{6in}{\includegraphics{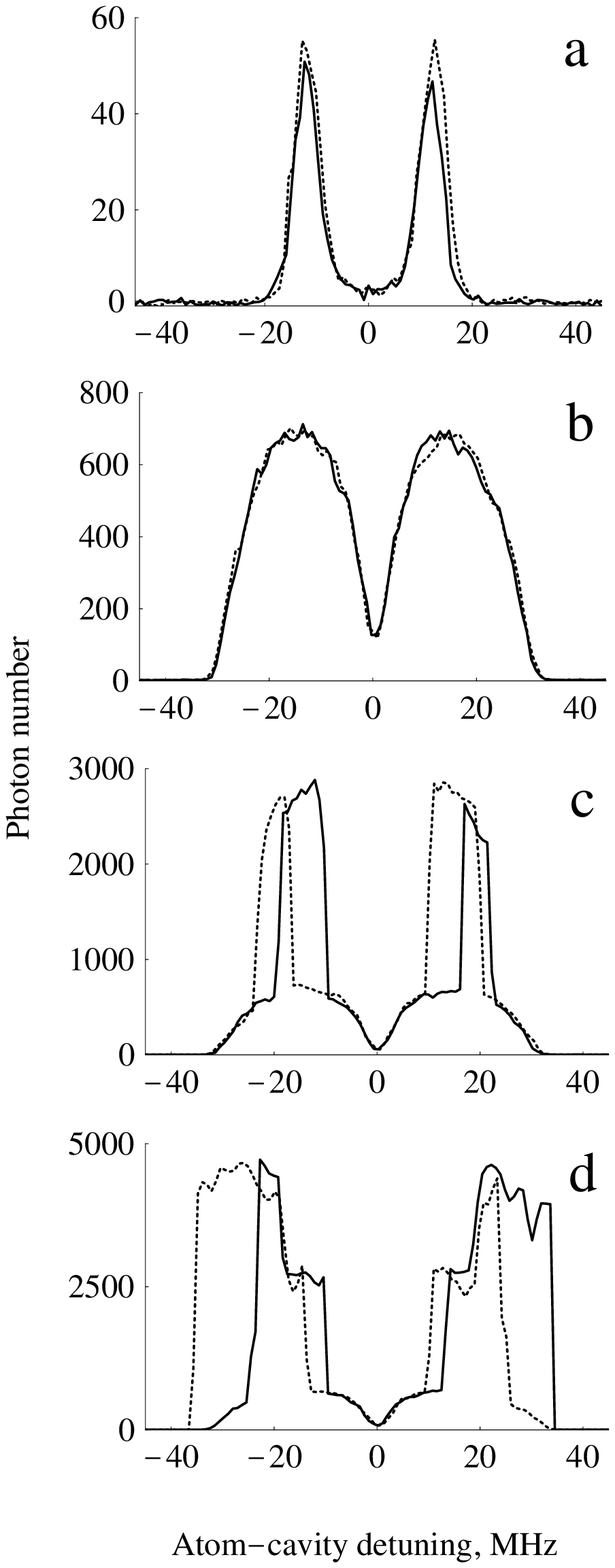}}\hspace{.5in}}
\caption{Cavity tuning lineshapes.  Estimated photon number $n$ vs.\
atom-cavity detuning $\Delta_{\rm cav}$.  Solid lines, positive
frequency scan (see text).  Dashed lines, negative frequency scan.
Estimated effective intracavity atoms: (a) \Neff = 18 , (b) $\Neff =
526$, (c) $\Neff = 657$, (d) $\Neff = 755$.}
\label{fig-lineshapes}
\efig

As density is increased, the two resonances broaden and increase in
amplitude.  For $\Neff \approx 630$ the second threshold appears as a
``spike'' near the peak of each resonance, and at $\Neff \approx 725$
a jump to a third branch appears.

The detuning lineshapes 
display strong asymmetry and hysteresis for atom numbers above the
second threshold.  
For densities in which sudden jumps occur, each traveling-wave
resonance of the cavity scanning lineshapes is highly asymmetric and
shows hysteresis as a function of scan direction.  This is most likely
due to interactions with the two Doppler-shifted fields at $\omega_a
\pm k v_0 \theta$.  For the general case in which neither field is
dominant, the atoms experience a complex bichromatic interaction with
these two fields.  A simple model has been shown to qualitatively
describe the lineshape asymmetry and hysteresis observed here.  More
detailed quantitative descriptions are in progress.

To examine the application of single-atom theory to our many-atom
microlaser, a semiclassical theory \cite{kwan-sc-theory} of the
microlaser analogous to the Lamb theory of the conventional laser
\cite{lamb-sc-theory-1964} has been developed.  In the case of
monochromatic field and zero atom-cavity detuning, the semiclassical
theory reduces to the rate equation model.  This provides a more
rigorous justification for Eqn.~\ref{eq-rate-equation}.  

%
%

We have also performed quantum trajectory simulations of a microlaser
with up to 5 intracavity atoms \cite{cfy-thesis}.  The many-atom
microlaser was shown to be in close agreement with the predictions
from a single-atom theory \cite{Filipowicz-PRA86}, with a perturbation
in the width of the photon number distribution due to cavity decay
during the interaction time.  

The connection between the microlaser with the conventional laser may
be illustrated by considering the effect of increased variation in $g
\tint$.  For simplicity suppose $f(g) = \delta(g-g_0)$ and interaction
times are distributed according to atom lifetime.  The dimensionless
gain is then
%
%
%
%
$P_{\rm emit} = {\tau^{-1}_a}\int_0^\infty d\tint \exp(-t/\tau_a) 
\sin^2(\sqrt{n+1} g \tint)
= {1/2}\left[
((n+1)g^2\tau_a^2)/(1+(n+1)g^2\tau_a^2)\right]$.  
With Eqn.~\ref{eq-rate-equation}, this expression describes a
conventional saturable gain laser, which is monostable and linear 
above threshold.

In the microlaser, for any finite degree of broadening in $g \tint$,
we have $P_{\rm emit}(n \rightarrow \infty) = 1/2$.  Therefore
conventional laser behavior is recovered in the limit of large photon
number.


In summary, the microlaser is shown to exhibit multistability due to
coherent atom-cavity interaction, which in conventional lasers is
hidden by spontaneous emission and other incoherent processes.

We have recently demonstrated that the microlaser field exhibits
sub-Poisson photon statistics \cite{wonshik}
even for hundreds of intracavity atoms.
Measurement of the spectrum of microlaser emission
\cite{Scully-linewidth} is also planned.

\begin{acknowledgments}
This research was supported by National Science Foundation grants
9876974-PHY and 0111370-CHE.  K.~An was supported by a Korea Research
Foundation Grant (KRF-2002-070-C00044).  The authors thank J.\ Thomas
of Duke University for assistance with the supersonic atom beam.
\end{acknowledgments}


\bibliography{biblio}

\end{document}